\documentclass[aps,reprint,onecolumn,notitlepage,11pt]{revtex4-1}
\usepackage{amsmath,amssymb,graphicx,subfigure,color}
\usepackage[hidelinks]{hyperref}

\begin{document}

% local institutional report number 
%\preprint{} 

\title{Geometric control of active collective motion}

\author{Maxime Theillard}
\author{Roberto Alonso-Matilla}
\author{David Saintillan}
\affiliation{Department of Mechanical and Aerospace Engineering, University of California San Diego, La Jolla, CA 92093, USA}

\date{\today}

\begin{abstract}
Recent experimental studies have shown that confinement can profoundly affect self-organization in semi-dilute active suspensions, leading to striking features such as the formation of steady and spontaneous vortices in circular domains and the emergence of unidirectional pumping motions in periodic racetrack geometries. Motivated by these findings, we analyze the two-dimensional dynamics in confined suspensions of active self-propelled swimmers using a mean-field kinetic theory where conservation equations for the particle configurations are coupled to the forced Navier-Stokes equations for the self-generated fluid flow. In circular domains, a systematic exploration of the parameter space casts light on three distinct states: equilibrium with no flow, stable vortex, and chaotic motion, and the transitions between these are explained and predicted quantitatively using a linearized theory. In periodic racetracks, similar transitions from equilibrium to net pumping to traveling waves to chaos are  observed in agreement with experimental observations and are also explained theoretically. Our results underscore the subtle effects of geometry on the morphology and dynamics of emerging patterns in active suspensions and pave the way for the control of active collective motion in microfluidic devices.
 \end{abstract}

\maketitle

\section{Introduction}

A common feature of many active matter systems is their ability to spontaneously self-organize into complex dynamic mesoscale structures above a certain density \cite{Marchetti13,Saintillan13,Saintillan15}. Such is the case of suspensions of motile bacteria \cite{Cisneros11,Wensink12,Dunkel13,Gachelin14}, cellular extracts \cite{Surrey01,Schaller10,Sanchez12}, collections of colloidal rollers \cite{Bricard13,Bricard15}, shaken grains \cite{Kudrolli08,Deseigne10}, among many others. Particle-particle interactions, whether long-ranged such as hydrodynamic or electrostatic interactions, or short-ranged such as direct contact forces, are the drivers of self-organization \cite{Marchetti13}. The symmetries of these interactions along with their coupling with system geometry dictates the structure and morphology of the emerging patterns, which include: steady vortices \cite{Bricard15,Tsang15}, asters \cite{Surrey01}, traveling bands \cite{Schaller10,Bricard13}, density shocks \cite{Lefauve14,Tsang16}, as well as more complex spatiotemporal chaotic patterns composed of unsteady jets and vortices \cite{Cisneros11}. 

Of interest to us in this work is the case of suspensions of hydrodynamically interacting slender self-propelled particles such as swimming bacteria \cite{Lauga09}. In these suspensions, particles exert dipolar stresses on the surrounding medium and also align in shear due to their elongated shape. The interplay between these two effects has been known to lead to hydrodynamic instabilities in the case of extensile particles or so-called pushers, which are thought to be responsible for the emergence of collective motion above a critical density \cite{Saintillan08,Saintillan08b,Subra09,Baskaran09}. In large unconfined systems, the collective dynamics in the nonlinear regime takes the form of unsteady chaotic motions reminiscent of high-Reynolds-number turbulence, characterized by strong jets and vortices, enhanced swimming speeds and diffusivities, and efficient fluid mixing \cite{Saintillan12}. 

Only recently have interactions with boundaries and dynamics in confined geometries gained attention in experiments. In dilute systems, it is well known that self-propelled particles accumulate at boundaries \cite{Berke08,Li09,Vladescu14,Figueroa15} as a result of both kinematic \cite{Li09,Elgeti13,Elgeti15,Ezhilan15,Ezhilan15b} and hydrodynamic mechanisms \cite{Berke08,Spagnolie12,Schaar15}. In complex geometries, transport of the particles along curved boundaries has also been exploited to design ratchets for concentrating microswimmers or directing their motion \cite{Galajda07,Hulme08,Lambert10,Altshuler13,Yariv14}. 
The case of semi-dilute to concentrated suspensions in confinement, however, has largely been unexplored but in a few studies. Wioland \textit{et al.} \cite{Wioland13} first analyzed the flow inside small droplets of a dense bacterial suspension squeezed between two flat plates. Rather than observing chaotic motion as in bulk systems, they reported the emergence of a steady vortex; detailed observation of the bacterial velocity field in fact revealed a more complex structure with a counter-rotating boundary layer surrounding the vortex core. This vortex was subsequently captured by Lushi \textit{et al.} \cite{Lushi14} in discrete particle simulations using a basic model accounting for dipolar hydrodynamic interactions as well as steric forces, where it was found that including hydrodynamic interactions is critical in order to correctly capture the counter-rotating boundary layer. Interactions between such vortices were also considered recently using a microfluidic lattice of circular chambers each containing one vortex and connected by junctions \cite{Wioland15}: in this case, hydrodynamic coupling was shown to produce synchronization on the scale of the lattice, with adjacent vortices rotating either in the same or opposite direction depending on the geometry of the junctions between chambers. 

The case of periodic geometries such as circular channels and racetracks has also been studied, where spontaneous flows have been reported in both bacterial \cite{Wioland16} and sperm \cite{Creppy15} suspensions above a critical density. In the case of bacteria, Wioland \textit{et al.} \cite{Wioland16} systematically studied the effect of geometry by varying the channel width. In very narrow channels, unidirectional flow takes place with a nearly parabolic velocity profile. Upon increasing channel width, flow patterns start exhibiting longitudinal oscillations leading to sinusoidal trajectories and eventually take the form of arrays of counter-rotating vortices. Longitudinal density waves were also reported in the case of dense semen \cite{Creppy15}. The observed transition to directed motion has been predicted in a number of models for active nematics \cite{Voituriez05,Ravnik13}, where extensile stresses were found to be the destabilizing factor leading to spontaneous flows. These models, however, neglected polarization, which plays an important role in setting the structure of the suspension in confined systems of  self-propelled particles \cite{Ezhilan15}; they also assumed anchoring boundary conditions for the nematic order parameter field at the channel boundaries, whereas the distribution of particle orientations near the walls appears to be dependent on flow conditions in experiments \cite{Wioland13,Wioland16}. A qualitative explanation for the transition can also be gleaned from recent studies on the effective rheology of active suspensions \cite{Gachelin13,Lopez15,Saintillan10,Alonso16}, where a decrease of the effective viscosity due to activity can lead at sufficiently high densities to a superfluid-like behavior in weak flows; this connection will be made clearer below. 

In this paper, we use numerical simulations based on a continuum kinetic model together with linear stability analyses to predict and characterize transitions to spontaneous flows and collective motion in various two-dimensional microfluidic geometries, with the aim of explaining the experimental observations discussed above. The governing equations are presented in section 2 and consist of evolution equations for the concentration, polarization and nematic order parameter, which are coupled to the Navier-Stokes equations for the mean-field flow induced by the swimmers. Results from simulations and theory are then discussed in section 3, where both circular domains and periodic racetracks are considered. We conclude in section 4.

\section{Model and simulation method}

\subsection{Continuum model}

We consider a collection of active Brownian particles suspended in a Newtonian fluid of density $\rho$ and shear viscosity $\mu$. The particles swim with velocity $V_{s}$ and have constant translational and rotational diffusivities $d_{t}$ and $d_{r}$, respectively. As a result of their self-propulsion, they also exert a net force dipole on the suspending fluid with stresslet strength $\sigma_{0}$, which we assume to be negative as is the case for extensile swimmers such as bacteria and sperm \cite{Hatwalne04,Drescher11}. The suspension, with mean number density $n$, is confined in a finite domain with characteristic dimension $H$, which will be defined more precisely later. Dimensional analysis of the governing equations identifies four relevant dimensionless groups:
\begin{equation}
Re=\frac{\rho H^{2} d_{r}}{\mu},  \quad Pe_s=\frac{V_{s}}{2d_{r}H},\quad \alpha = \frac{\sigma_{0} n}{\mu d_{r}},   \quad \Lambda = \frac{d_{r}d_{t}}{V_{s}^{2}}.
\end{equation}
The Reynolds number $Re$ is typically very small for active suspensions; it will be set to $10^{-6}$ in all the simulations and to zero in the stability results shown below. The swimming P\'eclet number $Pe_s$ denotes the ratio of the persistence length of swimmer trajectories to the size of the domain and is a measure of confinement. The activity parameter $\alpha$ compares the destabilizing effects of active stresses and of concentration to dissipative processes, namely~viscosity and orientation decorrelation by rotational diffusion. Finally, $\Lambda$ is a swimmer-specific parameter comparing diffusive processes to the strength of self-propulsion: the limit of $\Lambda\rightarrow 0$ describes athermal swimmers, whereas $\Lambda\rightarrow \infty$ corresponds to Brownian particles that do not swim. 

We adopt a two-dimensional continuum mean-field description of the active suspension based on the probability density function $\Psi(\mathbf{x},\mathbf{p},t)$ of finding a particle at position $\mathbf{x}$ with orientation $\mathbf{p}$ at time $t$, where $\mathbf{p}$ is a unit vector defining the swimming direction and orientation of the bacteria \cite{Saintillan08}. Following prior studies \cite{Saintillan13}, we approximate $\Psi$ in terms of its first three orientational moments: 
\begin{equation}
\Psi(\mathbf{x},\mathbf{p},t)\approx \frac{1}{2\pi}\left[c(\mathbf{x},t)+2 \mathbf{p}\cdot\mathbf{m}(\mathbf{x},t)+4\mathbf{pp}:\mathbf{D}(\mathbf{x},t)\right], \label{eq:psiexpansion} 
\end{equation}
where $c$, $\mathbf{m}$, and $\mathbf{D}$ are defined as integrals over the unit circle $C$ of orientations: 
\begin{align}
c(\mathbf{x},t)&=\int_{{C}}\Psi(\mathbf{x},\mathbf{p},t)\,d\mathbf{p}, \\
\mathbf{m}(\mathbf{x},t)&=\int_{{C}}\mathbf{p}\,\Psi(\mathbf{x},\mathbf{p},t)\,d\mathbf{p}, \\
\mathbf{D}(\mathbf{x},t)&=\int_{{C}}\left(\mathbf{pp}-\frac{\mathbf{I}}{2}\right)\Psi(\mathbf{x},\mathbf{p},t)\,d\mathbf{p}. 
\end{align}
The zeroth moment $c$ is the local concentration, whereas the first and second moments $\mathbf{m}$ and $\mathbf{D}$ describe the local polarization and nematic alignment in the suspension, respectively. Starting from a Smoluchowski equation for $\Psi(\mathbf{x},\mathbf{p},t)$ \cite{Saintillan08}, hierarchical evolution equations for the moments can be obtained, which are written as: 
\begin{align}
\partial_{t} c&=-\nabla\cdot\mathbf{F}_{c}, \label{eq:M12d}\\
\partial_{t} \mathbf{m}&=-\nabla\cdot\mathbf{F}_{m}+\tfrac{1}{2}\zeta \mathbf{E}\cdot\mathbf{m}-\mathbf{W}\cdot\mathbf{m}-\mathbf{m}, \label{eq:M22d}\\
\begin{split}
\partial_{t}\mathbf{D}&=-\nabla\cdot\mathbf{F}_{D}+\tfrac{1}{2}\zeta c\mathbf{E}+\tfrac{2}{3}\zeta\mathbf{E}\cdot\mathbf{D}-\tfrac{1}{3}\zeta \left(\mathbf{D}:\mathbf{E}\right)\mathbf{I} \\
&\quad+\mathbf{D}\cdot\mathbf{W}-\mathbf{W}\cdot\mathbf{D}-4\mathbf{D},
\end{split} \label{eq:M32d} 
\end{align}
where the dimensionless shape parameter $\zeta$ denotes Bretherton's constant \cite{Bretherton62}; we set $\zeta=1$ in the present study as is adequate for slender swimmers. Variables have been made dimensionless using length scale $H$ and time scale $d_r^{-1}$. The source terms on the right-hand side of Eqs.~(\ref{eq:M22d})--(\ref{eq:M32d})  arise from alignment and rotation by the rate-of-strain and vorticity tensors $\mathbf{E}$ and $\mathbf{W}$ of the disturbance velocity field $\mathbf{u}$,
and from rotational diffusion which promotes relaxation towards isotropy with $\mathbf{m}=\mathbf{0}$ and $\mathbf{D}=\mathbf{0}$.
The fluxes in the equations for $c$, $\mathbf{m}$, and $\mathbf{D}$ include contributions from advection by the flow, self-propulsion and translational diffusion, and are given by 
\begin{align}
\mathbf{F}_{c}&=\mathbf{u} \,c +2Pe_s \,\mathbf{m}-4\Lambda Pe_s^{2}\nabla c,  \label{eq:Fc} \\
 \mathbf{F}_{m}&=\mathbf{u}\,\mathbf{m}+2Pe_s\left(\mathbf{D}+c\tfrac{\mathbf{I}}{2}\right)-4\Lambda Pe_s^{2}\nabla\mathbf{m}, \label{eq:Fm}\\
\mathbf{F}_{D}&=\mathbf{u}\,\mathbf{D}+2Pe_s\left(\mathbf{T}-\mathbf{m}\tfrac{\mathbf{I}}{2}\right)-4\Lambda Pe_s^{2}\nabla \mathbf{D}, 
\end{align}
where the third-order tensor $\mathbf{T}$ is the third orientational moment and is related to the polarization according to the closure approximation implied by Eq.~(\ref{eq:psiexpansion}) as 
\begin{equation}
T_{ijk} =\frac{1}{4}\left(m_{i}\delta_{jk} +m_{j}\delta_{ik}+m_{k}\delta_{ij}\right). 
\end{equation}
Direct steric interactions between swimmers are neglected within this model. Their leading effect is expected to be an enhancement of local nematic alignment due to the elongated shape of the particles; this effect could easily be incorporated in our model using a nematic alignment potential as previously done by Ezhilan \textit{et al.} \cite{Ezhilan13}, though hydrodynamic interactions alone are sufficient to capture all the phenomenology observed in experiments. 

Finally, the disturbance fluid velocity field $\mathbf{u}$ satisfies the incompressible Navier-Stokes equations forced by the divergence of the active stress tensor $\alpha\mathbf{D}$:
\begin{align} 
\nabla\cdot\mathbf{u}=0, \quad 
Re \left(\partial_{t}\mathbf{u}+\mathbf{u}\cdot\nabla\mathbf{u}\right)=-\nabla p + \nabla^{2}\mathbf{u}+\alpha \nabla\cdot\mathbf{D}. 
\end{align}
Note that additional passive stresses also arise due to the inextensibility of the particles in the flow field they generate: we neglect those here as it can be shown that they only act to increase the Newtonian viscosity in the limit of weak flows relevant to the spontaneous flow transitions investigated here, and therefore only renormalize the value of $\alpha$ at the instability threshold. Including these stresses would be straightforward and the reader is referred to our previous work \cite{Saintillan10,Alonso16} on the rheology of active suspensions for more details. 

In all of our simulations and analysis, we enforce a no-slip boundary condition on the velocity $\mathbf{u}$ on the domain boundary $S$. The natural boundary condition for the particle distributions is a no-translational-flux condition on the probability density function $\Psi$, which translates into no-flux conditions on the orientational moments upon closure of the equations \cite{Ezhilan15}:
$\mathbf{n}\cdot\mathbf{F}_{c}=\mathbf{n}\cdot\mathbf{F}_{m}=\mathbf{n}\cdot\mathbf{F}_{D}=0,$
where $\mathbf{n}$ is the local unit normal. These conditions express the balance between self-propulsion and translational diffusion in the wall-normal direction, and were shown to correctly capture particle distributions near boundaries in confined systems \cite{Ezhilan15}.

\subsection{Numerical approach}

We solve the governing equations numerically using a hybrid finite-difference finite-volume framework \cite{Min06,Mirzadeh2011,Theillard13,Guittet15}. The method is implemented on adaptive quadtree grids and the domain boundaries are represented using the level-set method. At each time step of the algorithm, the moment equations (\ref{eq:M12d})--(\ref{eq:M32d}) are solved semi-implicitly: the diffusive terms are treated implicitly and the advective terms are computed using a semi-Lagrangian approach for improved stability, whereas the remaining coupling terms are treated explicitly. Values of the concentration field are stored at the cell centers to improve its total conservation, while the polarization and nematic order parameter fields are stored at the mesh nodes for better accuracy. Knowledge of the second moment $\mathbf{D}$ allows one to calculate the divergence of the active stress $\alpha \nabla \cdot \mathbf{D}$, which is an input to the Navier-Stokes solver \cite{Guittet15} used to update the fluid velocity field. A detailed description of the algorithm will be presented elsewhere.

\section{Results and discussion}

\subsection{Circular disks}

\begin{figure*}[t]
\centering
\includegraphics[scale=0.95]{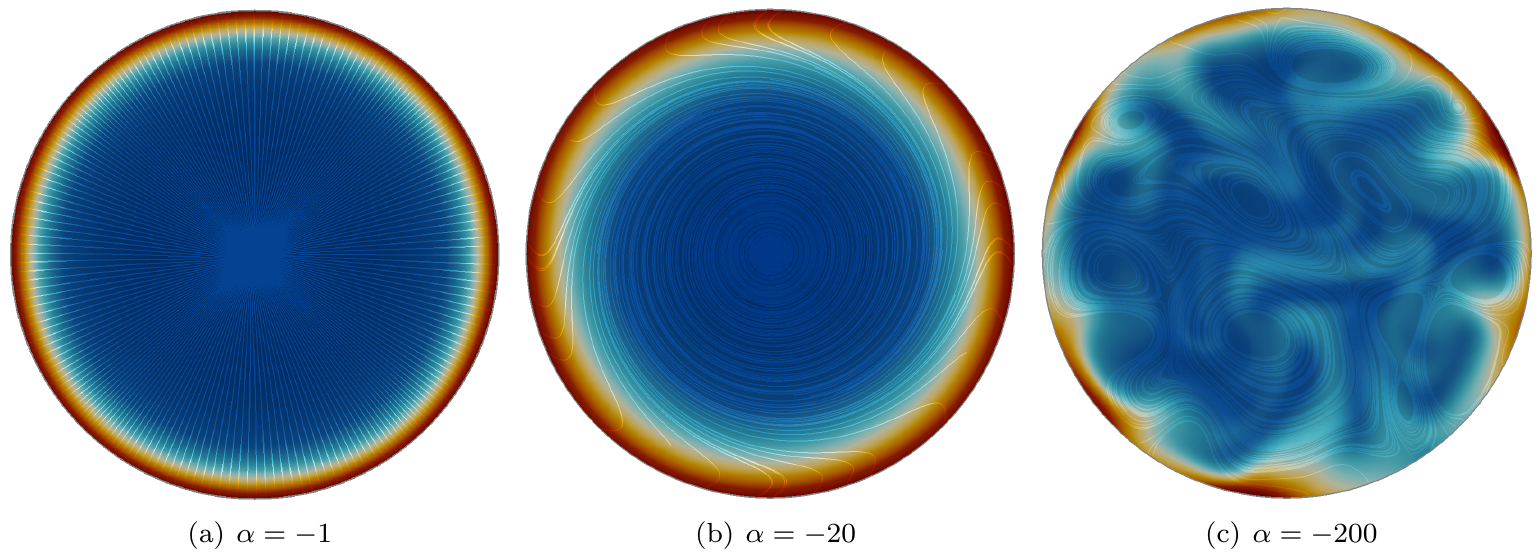}
\caption{Concentration profiles and streamlines of the net particle velocity $\mathbf{V}(\mathbf{x}) = 2Pe_{s}\mathbf{m}(\mathbf{x})/c(\mathbf{x}) + \mathbf{u}(\mathbf{x})$ showing three distinct regimes in a circular domain: (a) equilibrium base state with no flow (phase I); (b) double vortex flow (phase II); and (c) turbulent swirling state (phase III). Red indicates high concentration while blue is associated to a lower concentration. Results shown are for $Pe_s=0.5$ and $\Lambda=0.1$.  Also see electronic supplementary material for a movie showing the dynamics in each case. }
\label{fig:stream_plots}
\end{figure*}

\begin{figure}[t]
\centering
\includegraphics[width=0.45\textwidth]{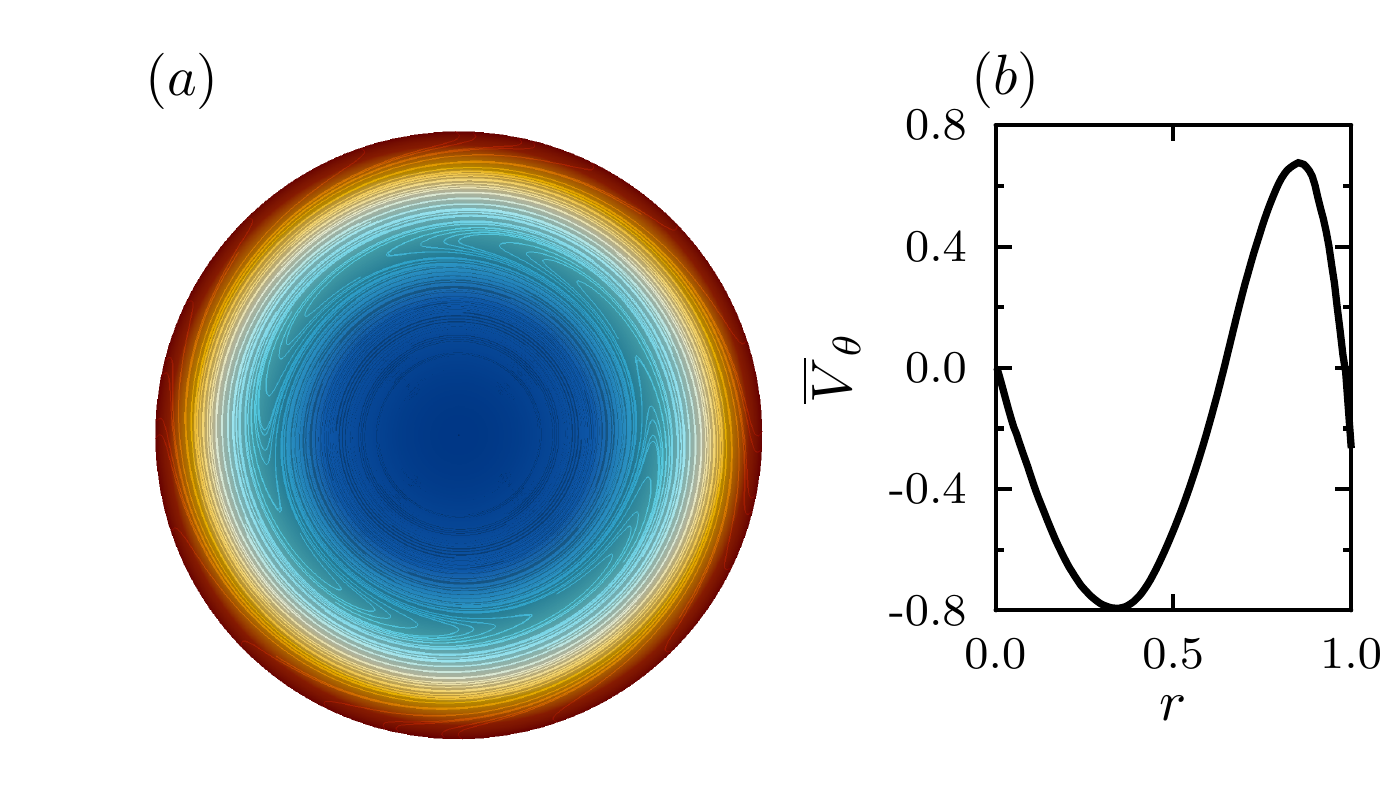}
\caption{Axisymmetric triple vortex in which the net particle velocity ${V_\theta}$ changes sign twice across the disk radius; this state is only rarely observed and therefore not included in the phase diagram of Fig.~\ref{fig:phasediagram}. (a) Concentration profile and streamlines of the net particle velocity. (b) Radial profile of the azimuthal particle velocity averaged over the azimuthal direction.  Parameter values for this simulation are $Pe_s=0.2$, $\alpha=-40$, and $\Lambda=0.2$.   }  
\label{fig:triplevortex}
\end{figure}

\begin{figure}[t]
\centering
\includegraphics[scale=0.9]{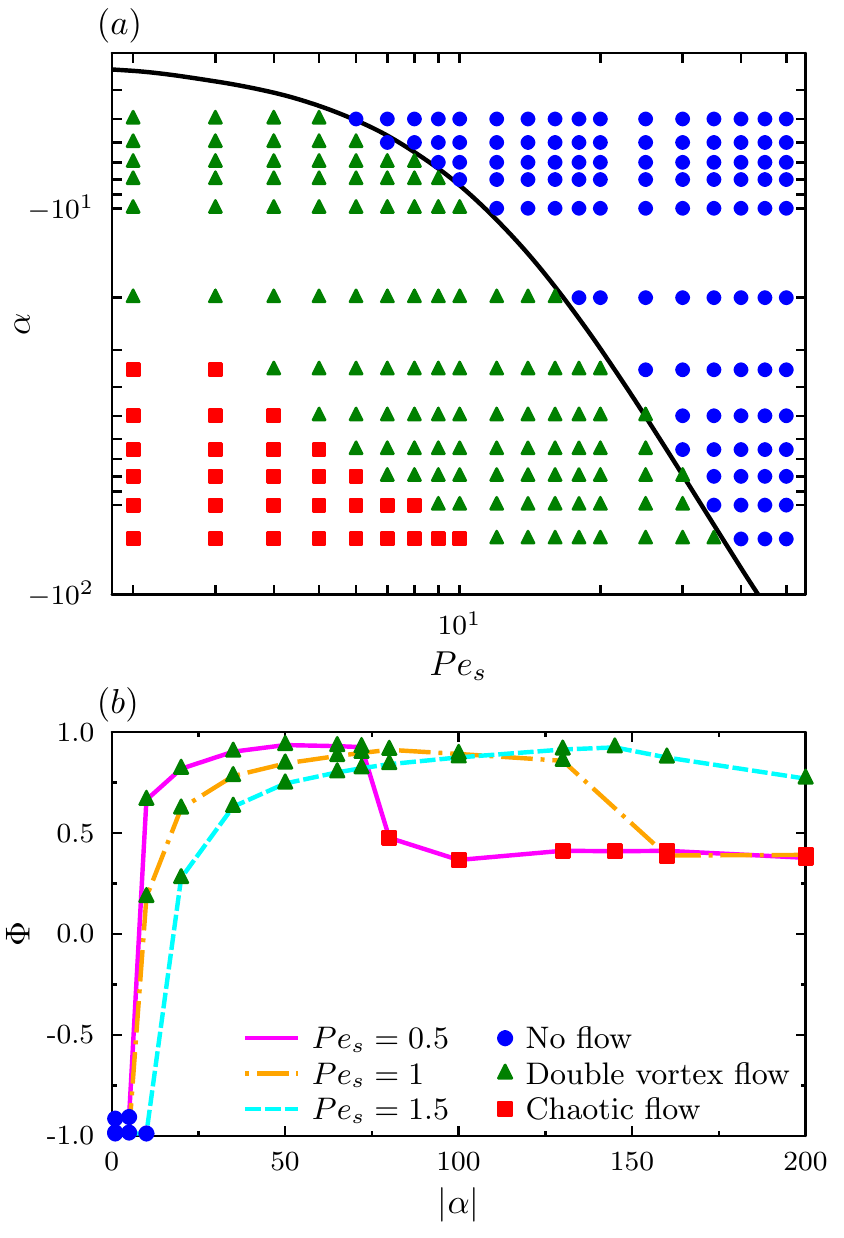}
\caption{Flow transitions in circular disks. (a) Phase diagram in the $(\alpha,Pe_s)$-plane for $\Lambda=0.1$ showing the transitions between phases I, II and III. The black curve shows the marginal stability for the equilibrium state of phase I as predicted by a linear stability analysis. (b) Order parameter $\Phi$ defined in Eq.~(\ref{eq:Phi}) as a function of activity $|\alpha|$ for three different values of $Pe_s$. $\Phi=-1$ corresponds to purely radial motion, $\Phi=+1$ to purely azimuthal motion, and $\Phi=\sqrt{2}-1$ to equal amounts of radial and azimuthal motion.   }  
\label{fig:phasediagram}
\end{figure}

Motivated by the experiments of Wioland \textit{et al.} \cite{Wioland13} in quasi-two-dimensional droplets, we first investigate the dynamics in circular domains, where we take the confinement length scale $H$ to be the radius of the disk. Our simulations in this case show that the collective self-organization depends critically on the level of activity (parameter $\alpha$) and degree of confinement (swimming P\'eclet number $Pe_s$). Specifically, three distinct phases illustrated in Fig.~\ref{fig:stream_plots} are observed depending on the values of $\alpha$ and $Pe_s$: an axisymmetric equilibrium state with no fluid flow (phase I), an axisymmetric and steady double vortex (phase II), and a turbulent-like unsteady chaotic state (phase III). In some cases, more complex axisymmetric flow patterns can also be observed before the transition to chaos, including triple vortices as illustrated in Fig.~\ref{fig:triplevortex}; such flow patterns are only very rarely observed and we do not discuss them further. Transitions between the three regimes, which are characterized in more detail below, can be captured in a phase diagram in the $(\alpha,Pe_s)$-plane as shown in Fig.~\ref{fig:phasediagram}(a), where we find that either increasing activity or decreasing confinement successively destabilizes phase I into phase II followed by phase III. 
While distinguishing between these states is straightforward by simple observation of the dynamics, we also introduce an order parameter as a quantitative measure: 
\begin{equation}
\Phi=\Big\langle \frac{2}{\pi}\int_{\Omega}\frac{|V_{\theta}|}{|\mathbf{V}|}d\mathbf{x}-1\Big\rangle_{t}, \label{eq:Phi}
\end{equation}
where $\mathbf{V}=2Pe_s\, \mathbf{m}(\mathbf{x})/c(\mathbf{x})+\mathbf{u}(\mathbf{x})$ is the net velocity of the active particles due to both swimming and advection by the flow, $V_\theta$ is its azimuthal component, $\Omega$ is the circular domain, and $\langle\cdot\rangle_t$ denotes a time average. Note that the velocity $\mathbf{V}$ is the same as that measured in experiments, which typically perform particle-image velocimetry based on swimmer displacements \cite{Wioland13,Wioland16}. The order parameter in Eq.~(\ref{eq:Phi}) is defined such that $\Phi=-1$ for purely radial motion (phase I), $\Phi=+1$ for purely azimuthal motion (phase II), and $\Phi=\sqrt{2}-1$ for a system in which motion occurs equally along the radial and azimuthal directions (phase III). A plot of $\Phi$ vs $|\alpha|$ for various values of $Pe_s$ is shown in Fig.~\ref{fig:phasediagram}(b), where it indeed jumps from $-1$ to $\approx +1$ as the transition from phase I to phase II occurs, before eventually decreasing to a value close to $\sqrt{2}-1$ as the chaotic state of phase III emerges. We now proceed to characterize the three phases in more detail. 

\vspace{0.3cm}

\noindent \textit{Phase I: Equilibrium state with no flow. ---} This regime, illustrated in Fig.~\ref{fig:stream_plots}(a), occurs for low levels of activity (small $|\alpha|$) and strong levels of confinement (large $Pe_s$), and is characterized by the absence of hydrodynamic flow. As is known to be the case in dilute confined active suspensions \cite{Ezhilan15}, particles tend to accumulate near the system boundaries and on average point towards the boundary, which leads to a net radial polarization $m_{r}(r)>0$ that reaches its maximum at the walls. The azimuthal polarization $m_{\theta}$ is zero in this case, as are off-diagonal components $D_{r\theta}$ and $D_{\theta r}$ of the nematic order tensor. An analytical solution for the first two moments can in fact be derived in this case by neglecting nematic alignment ($\mathbf{D}=\mathbf{0}$), which is a good approximation as shown by full numerical simulations. The solution, given in Appendix A, predicts a dimensionless characteristic thickness for the wall accumulation layer given by
\begin{equation}
\Omega^{-1}=2\Lambda Pe_s \sqrt{2/(1+2\Lambda)}, 
\end{equation}
which can also be interpreted as the length scale over which the effects of the boundary are screened by diffusive processes \cite{Yan15}. As first discussed by Ezhilan \& Saintillan \cite{Ezhilan15}, two interesting limiting cases are found. When $\Lambda\rightarrow 0$ (strong-propulsion limit), the thickness of the layer is set by the balance of swimming and translational diffusion and is given by $\Omega^{-1}H\approx \sqrt{2}\, d_t/V_s$;  on the other hand, the weak-propulsion limit of  $\Lambda\rightarrow \infty$ yields a  thickness of $\Omega^{-1}H \approx \sqrt{d_t/d_r}$, which is a purely diffusive length scale. 

\begin{figure}[t]
\centering
\includegraphics[scale=0.9]{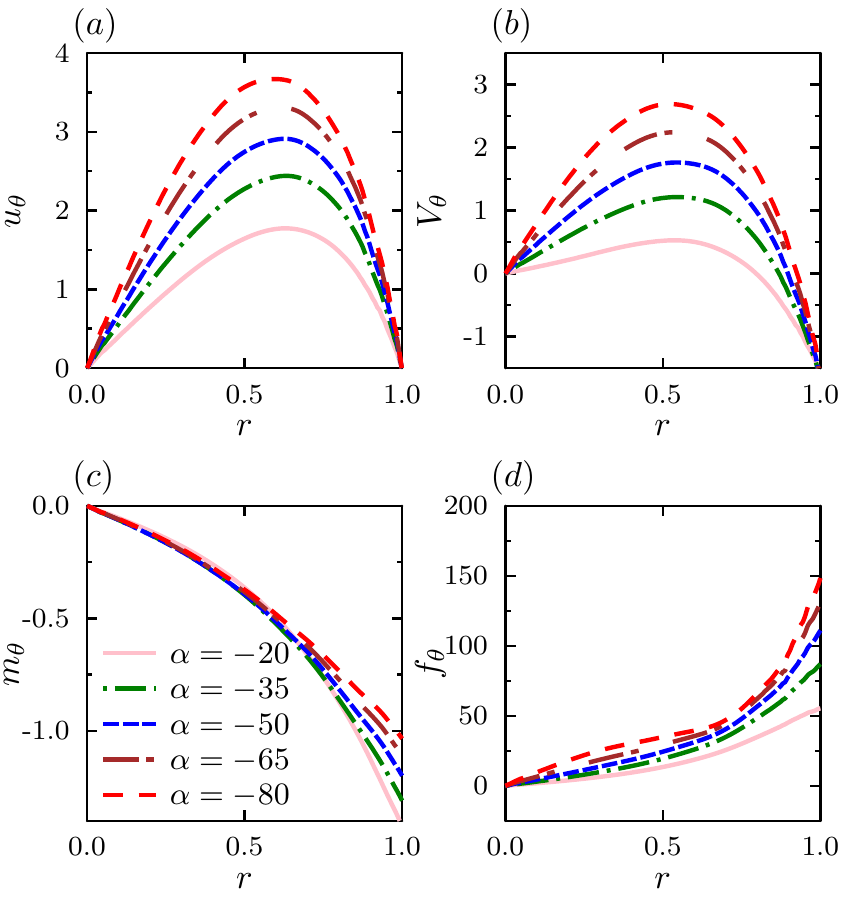}
\caption{Double vortex flow in a circular disk (phase II): radial profiles of (a) the azimuthal fluid velocity $u_\theta$, (b) the azimuthal net particle velocity $V_\theta$, (c) the azimuthal polarization $m_\theta$, and (d) the active flow forcing $f_{\theta}=\alpha r^{-2} \partial_{r}(r^2 D_{r\theta})$ for different levels of activity, for $Pe_s=1$ and $\Lambda=0.1$.   }  
\label{fig:profiles}
\end{figure}

\begin{figure}[t]
\centering
\includegraphics[scale=0.9]{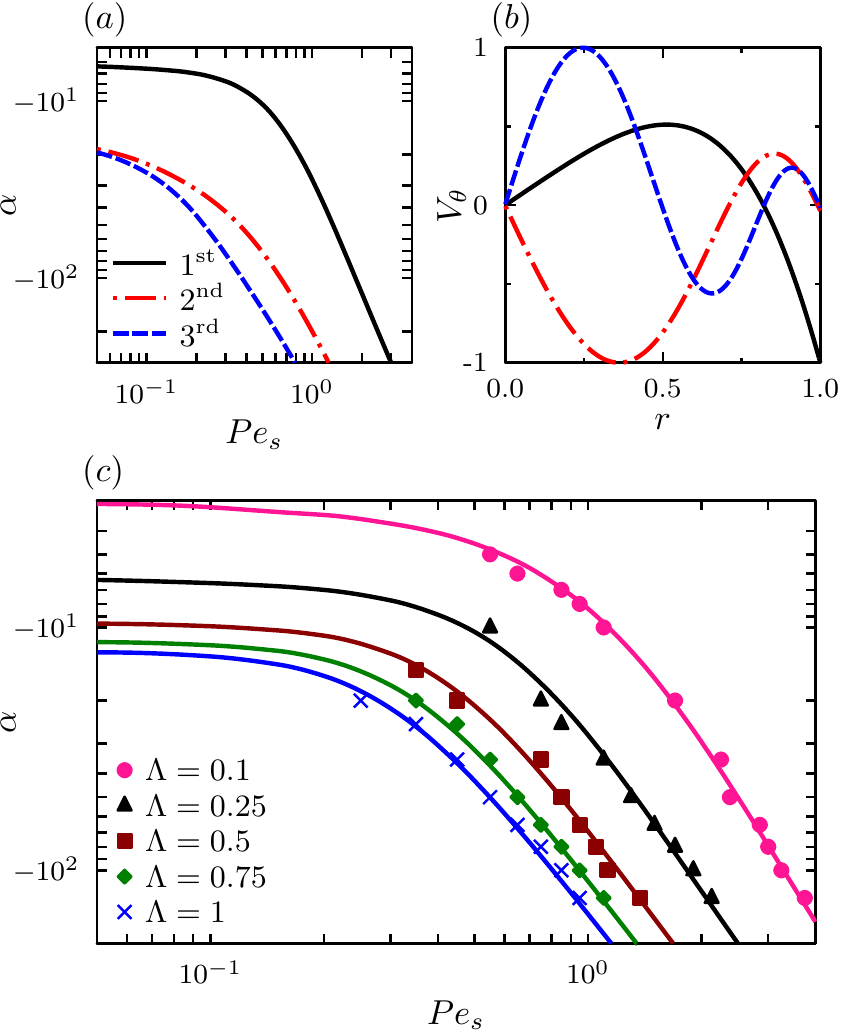}
\caption{Linear stability of the equilibrium base state (phase I) in a circular disk. (a) Marginal stability curves in the $(\alpha,Pe_s)$-plane for the first three unstable modes at $\Lambda=0.25$. (b) Unstable eigenmodes for the net particle velocity $V_\theta$ for $Pe_s=0.5$, and $\Lambda=0.25$ at the onset of instability. (c) Marginal stability curves for the first unstable mode for different values of $\Lambda$ obtained by linear stability analysis (full lines); symbols show the marginal curve for the transition from phase I to phase II in numerical simulations.   }  
\label{fig:stabilitydisk}
\end{figure}

\vspace{0.3cm}

\noindent \textit{Phase II: Steady axisymmetric vortex. ---}  Upon decreasing confinement or increasing the level of activity, a first transition is observed from the equilibrium state to a steady axisymmetric double vortex shown in Fig.~\ref{fig:stream_plots}(b), as a consequence of the coupled effect of nematic alignment induced by hydrodynamic interactions and base-state heterogeneities in the concentration and radial polarization profiles. The direction of rotation in this case is arbitrary with equal probabilities for clockwise and counter-clockwise motions, though it is found to remain the same for the duration of the simulation. More details on this regime, which is identical to that reported in the experiments of Wioland \textit{et al.} \cite{Wioland13}, are shown in Fig.~\ref{fig:profiles}, where profiles of the azimuthal components of the fluid velocity $u_\theta$, net particle velocity $V_\theta$, polarization $m_\theta$, and active flow forcing $f_{\theta}=\alpha r^{-2} \partial_{r}(r^2 D_{r\theta})$ are plotted for different parameter values. While the azimuthal fluid velocity $u_{\theta}$ always points in the same direction, the net particle velocity $V_\theta$ follows the fluid flow near the center of the domain but changes sign at a distance away from the boundary, indicating that particles near the circular wall move against the local flow in agreement with experiments and previous models \cite{Wioland13,Lushi14}. Fig.~\ref{fig:profiles}(c) in fact shows that the azimuthal polarization is negative, i.e.~particles swim against the flow everywhere but only overcome it near the wall.  This observation is consistent with the well-known phenomenon of upstream swimming, by which near-wall bacteria tend to swim against the fluid in pressure-driven channel flows of active suspensions \cite{Hill07,Kaya12,Kantsler14,Ezhilan15,Mathijssen16}. As the transition from phase I to phase II occurs, we also find that particle accumulation at the boundary is weakened, which is a consequence of the alignment of the particles with the hydrodynamic flow, which reduces the wall-normal polarization and therefore hinders the ability of the particles to swim towards the wall; a similar effect is again also known to occur in pressure-driven channel flows \cite{Figueroa15,Ezhilan15}.

Mechanistic insight into the formation of the double vortex can be gained by performing a linear stability analysis, in which we theoretically analyze the growth of axisymmetric perturbations to the equilibrium base state with no flow (phase I). Results from this analysis are summarized in Fig.~\ref{fig:stabilitydisk}. The analysis indeed reveals a linear instability of the equilibrium base state, with a hierarchy of unstable modes for which we plot the marginal stability curves in the $(\alpha,Pe_s)$-plane in Fig.~\ref{fig:stabilitydisk}(a) and the net azimuthal velocity $V_\theta$ in Fig.~\ref{fig:stabilitydisk}(b). The first unstable mode has a structure that is very similar to the nonlinear flow field of Fig.~\ref{fig:profiles}(b), with a vortex core and a counter-rotating boundary layer near the domain wall. Subsequent modes, which become unstable at higher values of $|\alpha|$ as shown in Fig.~\ref{fig:stabilitydisk}(a), exhibit more and more complex structures with alternating layers rotating in opposite directions. These modes are only very rarely observed in simulations, and an example of a triple vortex similar for the second unstable mode is shown in Fig.~\ref{fig:triplevortex}; in most cases, however, we find that the double vortex instead destabilizes directly into the chaotic state.

The linearized theory also sheds light on the mechanism for the transition, which can be summarized as follows: (i) At equilibrium (phase I), particles near the boundary have a net polarization towards the wall \cite{Ezhilan15}; (ii) A weak azimuthal flow perturbation causes these particles to rotate due to shear and align at an angle with respect to the radial direction, leading to a net shear nematic alignment $D_{r\theta}$ which is strongest near the wall, as well as an azimuthal polarization in the direction opposite to the fluid flow; (iii) Hydrodynamic disturbances induced by the swimming activity in the nematically-aligned region tend to reinforce the flow perturbation in the case of pushers via the active forcing term $f_{\theta}=\alpha r^{-2} \partial_{r}(r^2 D_{r\theta})$ in the $\theta$-momentum equation. In particular, it can be checked both in theory and simulations that suspensions of pullers for which $\alpha>0$ are always stable and only exhibit phase I.  A more quantitative comparison between theory and simulations is shown in Fig.~\ref{fig:stabilitydisk}(c), where the marginal stability curves for the first unstable mode for different values of $\Lambda$ are found to match the numerical transition from phase I to phase II. Fig.~\ref{fig:stabilitydisk}(c) also shows the influence of the parameter $\Lambda$: increasing its value has a stabilizing effect due to diffusion, which smoothes out the wall accumulation layer responsible for driving the flow. 

\begin{figure}[t]
\centering
\includegraphics[scale=0.9]{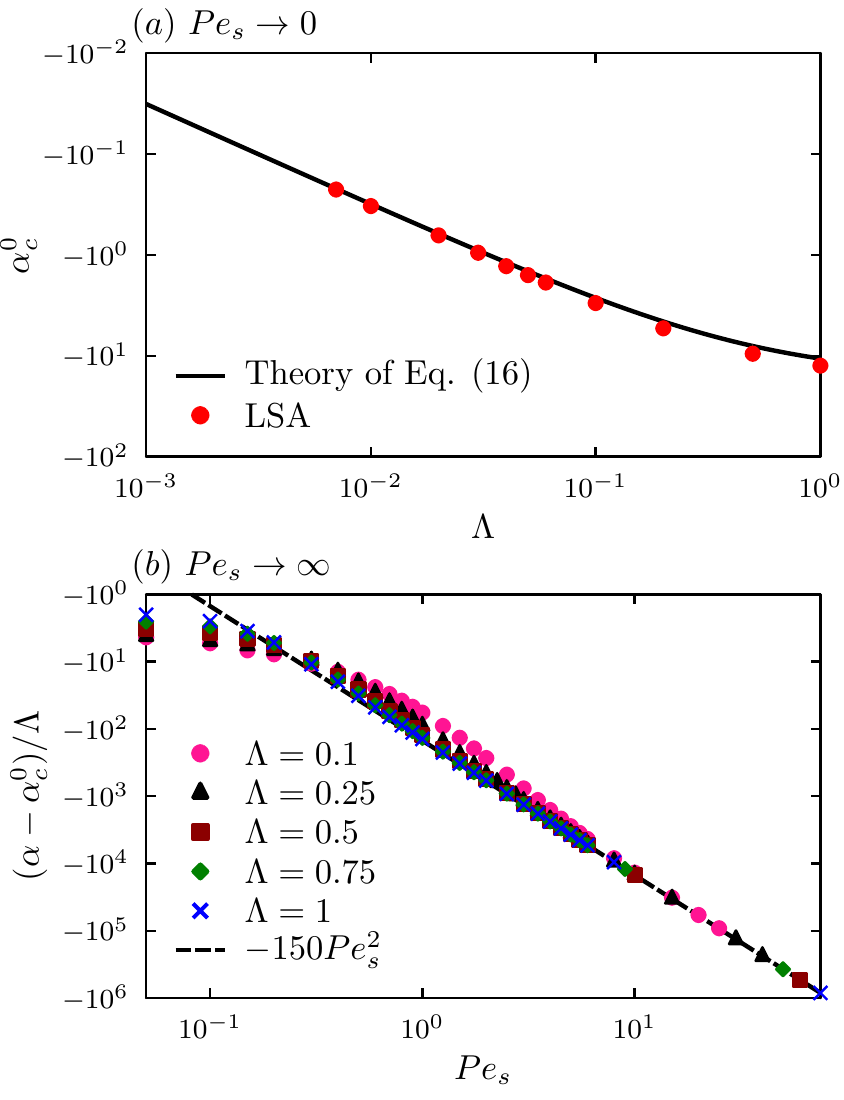}
\caption{Analytical prediction of the marginal stability curve. (a) Low-$Pe_s$ asymptote $\alpha^0_c$ for the critical value of the activity parameter leading to instability of the equilibrium state: the plot compares the theoretical prediction of Eq.~(\ref{eq:lowPe}) to the full numerical solution of the linear stability problem (LSA). (b) High-$Pe_s$ asymptote: the plot shows $(\alpha-\alpha^0_c)/\Lambda$ as a function of $Pe_s$ and confirms the scaling prediction of Eq.~(\ref{eq:highPe}) with fitting parameter $\beta\approx 150$.  }  
\label{fig:stabilitydisk2}
\end{figure}

The low- and high-P\'eclet limits for the marginal stability of the equilibrium state can also be characterized more precisely as illustrated in Fig.~\ref{fig:stabilitydisk2}. In weakly confined systems ($Pe_s\rightarrow 0$), the marginal value of $\alpha$ for instability becomes independent of system size as shown by the constant asymptote in Fig.~\ref{fig:stabilitydisk}(c): in this case, the instability is primarily driven by processes inside the accumulation layer and an approximate expression for the critical value of $\alpha$ is derived in Appendix B as 
\begin{equation}
\alpha^0_c \approx -\frac{32\Lambda}{1+2\Lambda} \qquad \mbox{as}\quad Pe_s\rightarrow 0, \label{eq:lowPe} 
\end{equation}
which matches the numerical solution of the linear stability problem excellently, especially for small values of $\Lambda$, as shown in Fig.~\ref{fig:stabilitydisk2}(a). In the limit of strong confinement ($Pe_s\rightarrow \infty$), we expect translational diffusion to be the main stabilizing factor, which suggests an asymptote of the form
\begin{equation}
\alpha^\infty_c\approx -\beta \Lambda Pe_s^2 \qquad \mbox{as}\quad Pe_s\rightarrow \infty, \label{eq:highPe} 
\end{equation}
where $\beta$ is an unknown constant. Plotting $(\alpha-\alpha^0_c)/\Lambda$ as a function of $Pe_s$ in Fig.~\ref{fig:stabilitydisk2}(b) indeed collapses all the marginal stability curves onto a single power-law consistent with Eq.~(\ref{eq:highPe}), where the fitting parameter $\beta$ is found to depend very weakly on $\Lambda$ and to asymptote to $\beta\approx 150$ at high values of $\Lambda$. A composite approximation for the marginal stability curve can therefore be written as 
\begin{equation}
\alpha_c \approx -\frac{32\Lambda}{1+2\Lambda}-150 \Lambda Pe_s^2, 
\end{equation} 
which provides an excellent fit to our numerical data over a wide range of parameter values. The high-$Pe_s$ asymptote of Eq.~(\ref{eq:highPe}) can also be used to define a critical domain diameter for the emergence a double vortex in strong confinement: 
\begin{equation}
D_c\approx 5\sqrt{\frac{6\mu d_t}{n\sigma_0}}. 
\end{equation}
Interestingly, this critical domain size only depends upon $d_t$ and does not involve $d_r$. The scaling with number density $D_c\sim n^{-1/2}$ also differs from the scaling of $n^{-1}$ for the critical system size for the onset of collective motion in bulk systems \cite{Hohenegger10}.

\vspace{0.3cm}

\noindent \textit{Phase III: Turbulent swirling state. ---}  As the level of activity keeps increasing and confinement is decreased, phase II becomes unstable itself leading to phase III, which is an unsteady chaotic state analogous to that observed in unbounded systems \cite{Saintillan08b,Ezhilan13}. There is no clear structure to the flow in this case, which is instead characterized by local jets and vortices driven by active stresses.  Note that the transition to phase III is not predicted by our linear theory, which can only account for axisymmetric disturbances. Based on previous analyses of unbounded active turbulence \cite{Saintillan08b,Subra09,Baskaran09}, we hypothesize that the transition nonetheless results from a linear instability of the double vortex of phase II to two-dimensional disturbances, though a more detailed theoretical analysis remains to be done in this case.

\begin{figure*}[t]
\centering
\includegraphics[scale=1]{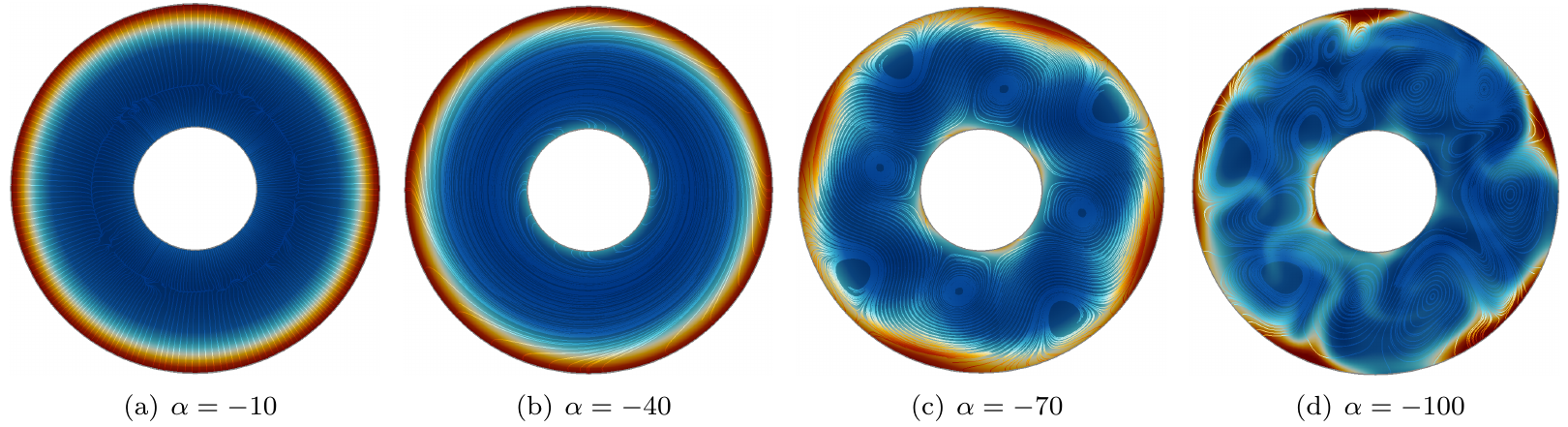}
\caption{Concentration profiles and streamlines of the net particle velocity showing four distinct regimes in a circular annulus: (a) equilibrium base state with no flow (phase I); (b) axisymmetric azimuthal flow with net pumping (phase II); (c) azimuthal flow with net pumping and traveling waves (phase III); and (d) turbulent swirling state (phase IV). Red indicates high concentration while blue is associated to a lower concentration. Results shown are for $Pe_s=0.5$ and $\Lambda=0.5$.  Also see electronic supplementary material for a movie showing the dynamics in each case.  }  
\label{fig:annulus}
\end{figure*}

\subsection{Periodic channels: circular annuli}

\begin{figure}[t]
\centering
\includegraphics[scale=0.9]{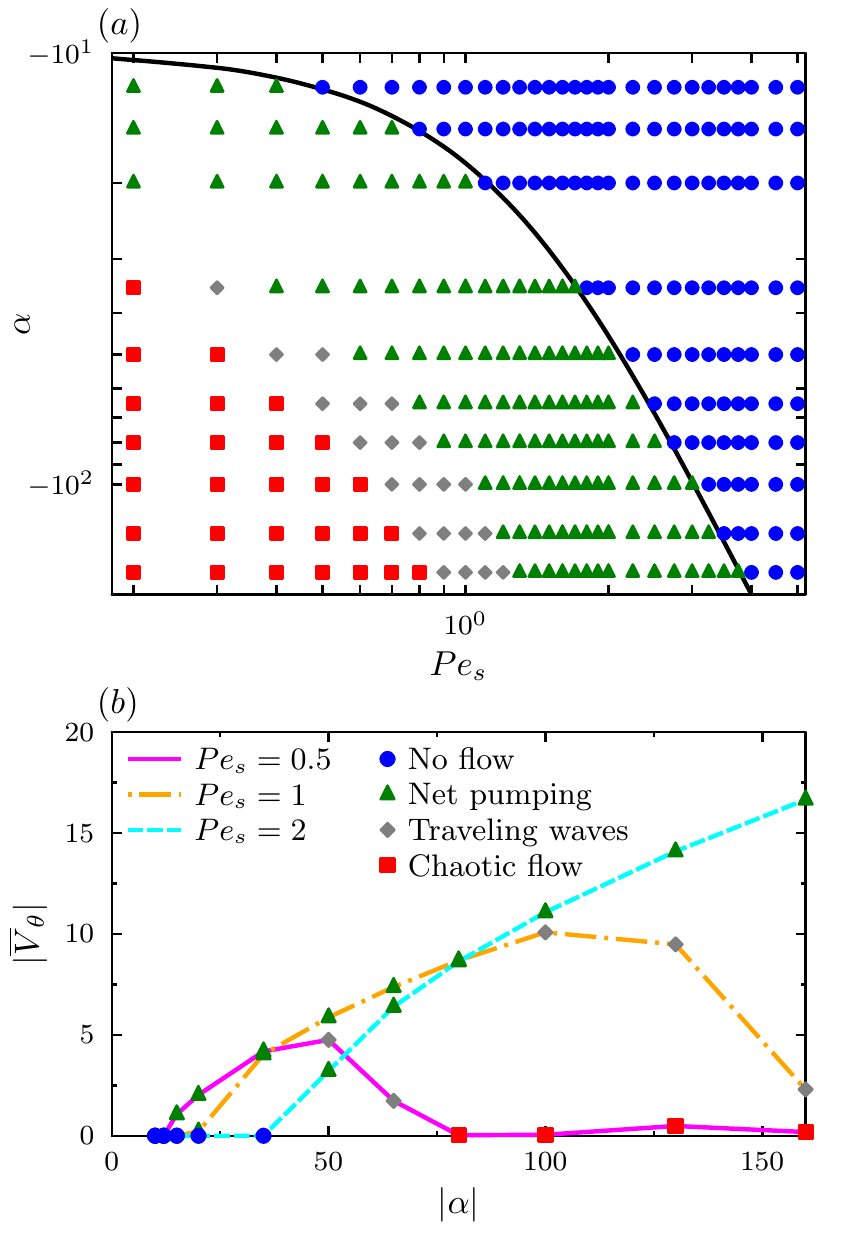}
\caption{Flow transitions in circular annuli. (a) Phase diagram in the $(\alpha,Pe_s)$ plane for $\Lambda=0.5$ and $r_\mathrm{min}=1$, showing the transitions between phases I, II, III, and IV. The black curve shows the marginal stability for the equilibrium state of phase I as predicted by a linear stability analysis. (b) Mean azimuthal particle velocity $|\overline{V}_\theta|$ as a function of activity $|\alpha|$ for three different values of $Pe_s$. }  
\label{fig:phasediagannulus}
\end{figure}

We now turn our attention to the case of periodic channels and first focus on annulus geometries in which the two boundaries are concentric circles. We use the channel halfwidth as the characteristic length scale $H$, and introduce an additional parameter as the dimensionless inner radius $r_{\mathrm{min}}$. The phenomenology in this case is illustrated in Fig.~\ref{fig:annulus}, where four distinct regimes are observed: an axisymmetric equilibrium state with no fluid flow (phase I), an axisymmetric spontaneously flowing state with net fluid pumping (phase II), a spontaneously flowing state with net fluid pumping and traveling density waves (phase III), and a turbulent-like chaotic state (phase IV). With the exception of phase III (traveling waves), these regimes are qualitatively similar to those found in circular disks. Transitions  between the different states also show similar trends in the $(\alpha,Pe_s)$-plane, as depicted in the phase diagram of Fig.~\ref{fig:phasediagannulus}(a) where phase III occupies a thin region between phases II and IV. Transitions are also characterized in Fig.~\ref{fig:phasediagannulus}(b), where the absolute value of the mean azimuthal particle velocity $|\overline{V}_\theta|$ is plotted as a function of $|\alpha|$ for different P\'eclet numbers. As expected, the transition from equilibrium to net pumping is accompanied by a bifurcation from zero to a finite flow rate; the flow rate initially increases with activity up to the point when traveling waves appear, after which it start decreasing to reach zero in the chaotic state. We now discuss the various regimes in more detail. 

\begin{figure}[t]
\centering
\includegraphics[scale=0.9]{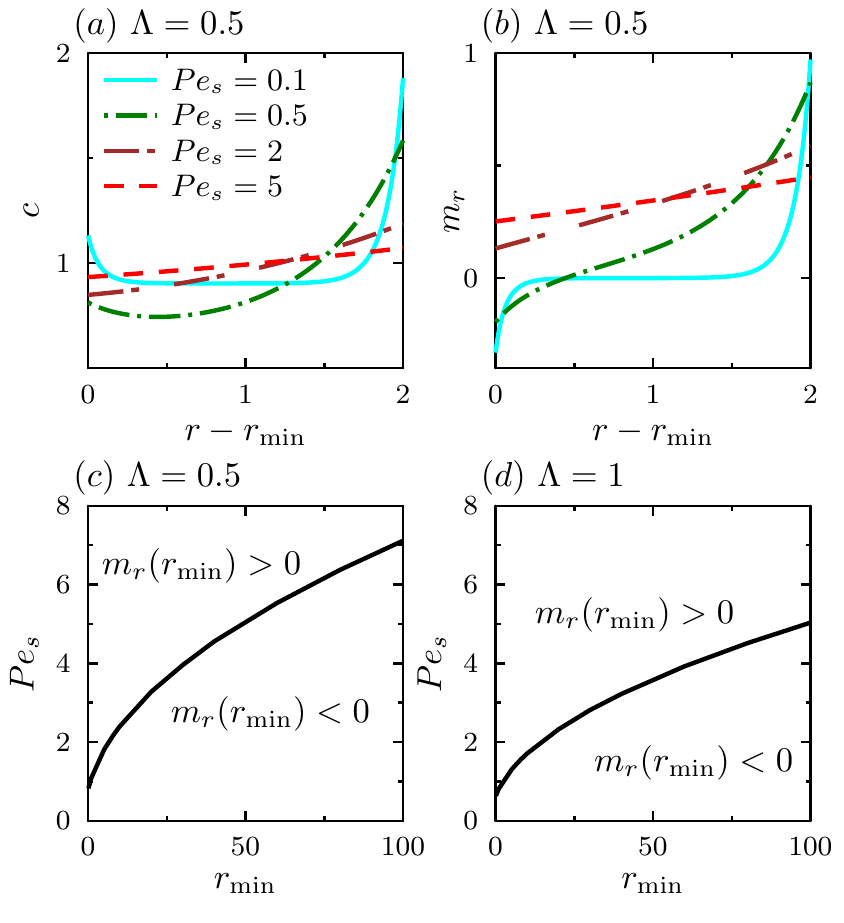}
\caption{Equilibrium distribution (phase I) inside an annulus. (a) Concentration and (b) polarization profiles across the gap as functions of $Pe_s$, for $\Lambda=0.5$ and $r_\mathrm{min}=Pe_s$, which is equivalent to varying gap width in dimensional terms. (c)-(d) Transition from accumulation to depletion at $r=r_\mathrm{min}$ in the $(Pe_s,r_\mathrm{min})$-plane for $\Lambda=0.5$ and $1$, respectively. }  
\label{fig:basestateannulus}
\end{figure}

\vspace{0.3cm}

\noindent \textit{Phase I: Equilibrium state with no flow. ---} The equilibrium state shown in Fig.~\ref{fig:annulus}(a) is very similar to that found in circular domains, and is characterized by particle accumulation and wall normal polarization at the outer boundary. The particle distribution at the inner boundary, however, shows a subtle dependence on parameter values and can either show accumulation (with $m_r<0$) or depletion (with $m_r>0$). This is illustrated in Fig.~\ref{fig:basestateannulus}(a)-(b), where we plot radial concentration and polarization profiles across the gap for different channel widths (i.e. different values of $Pe_s$ and $r_\mathrm{min}$ at a fixed ratio of $Pe_s/r_\mathrm{min}$): in wide channels (small $Pe_s$) accumulation occurs at both boundaries, but a depletion is observed in narrow channels (large $Pe_s$) in which case attraction by the outer boundary dominates due to curvature effects and propagates across the entire gap due to diffusion in spite of the presence of the inner wall. Depletion at the inner boundary occurs in strongly diffusive systems or under strong confinement, where both the concentration and polarization profiles become linear across the gap. The transition between the two types of distributions is captured in Fig.~\ref{fig:basestateannulus}(c) in terms of $Pe_s$ and $r_\mathrm{min}$ for $\Lambda=0.5$:  as $r_\mathrm{min}$ decreases for a fixed gap width, the curvature of the outer boundary becomes more positive while that of the inner boundary becomes more negative, leading to a higher symmetry breaking between walls thus fostering depletion at the inner wall. The same transition is captured in Fig.~\ref{fig:basestateannulus}(d) for $\Lambda=1$, where particle attraction by the outer wall is yet stronger as translational diffusion is enhanced.

\begin{figure}[t]
\centering
\includegraphics[scale=0.85]{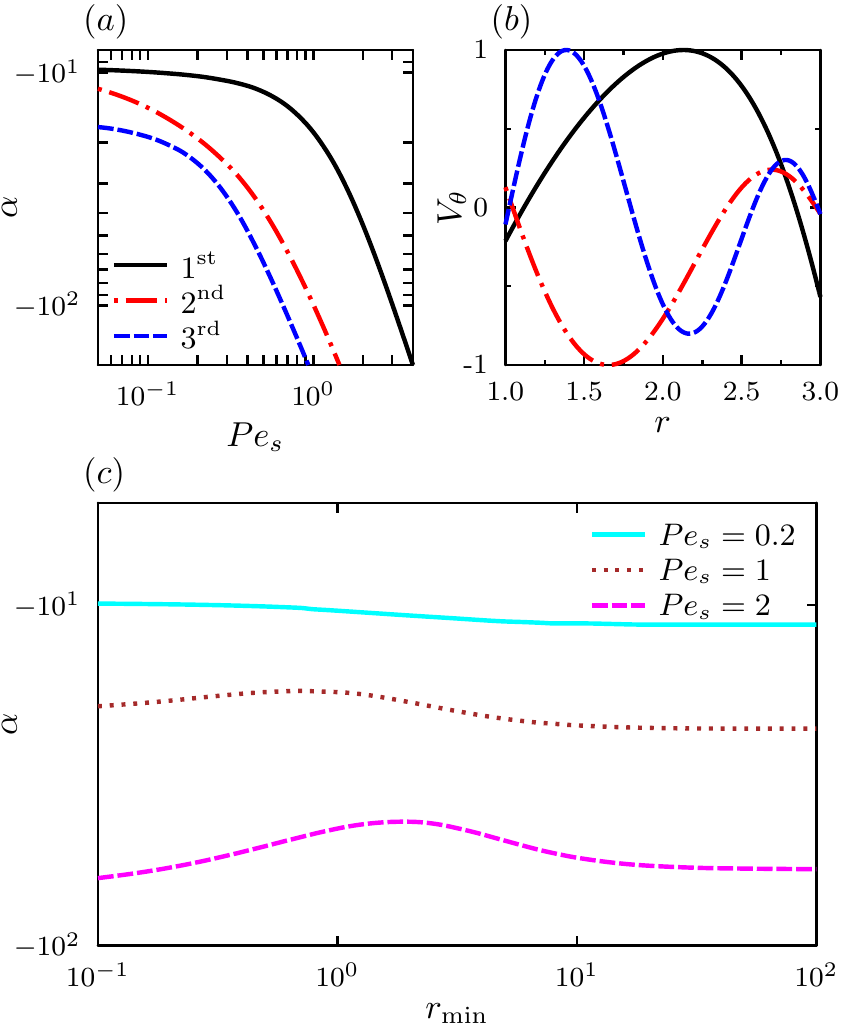}
\caption{Linear stability of the equilibrium base state (phase I) in a circular annulus. (a) Marginal stability curves in the $(\alpha,Pe)$-plane for the first unstable modes at $\Lambda=0.5$ and $r_{\mathrm{min}}=1$ (b) Unstable eigenmodes for the net particle velocity $V_\theta$ for $Pe=0.5$, $\Lambda=0.5$, and $r_\mathrm{min}=1$ at the onset of instability. (c) Marginal stability curves for the first unstable mode in the $(\alpha,r_\mathrm{min})$ plane for different values of $Pe_s$ and for $\Lambda=0.5$.   }  
\label{fig:annulusstability}
\end{figure}

\vspace{0.3cm}

\noindent \textit{Phase II: Spontaneous flow with net pumping. ---} As activity is increased or confinement is decreased, a first transition to an axisymmetric flowing state with net fluid pumping occurs (phase II). As previously seen in Fig.~\ref{fig:phasediagannulus}(b), the mean azimuthal velocity $|\overline{V}_\theta|$ is non-zero in this regime and increases monotonically with the level of activity up to the point where traveling waves appear (phase III below). The transition to unidirectional flow is similar to that reported in both bacterial \cite{Wioland16} and sperm \cite{Creppy15} suspensions. It was also predicted in a few previous theoretical and numerical models \cite{Voituriez05,Ravnik13}, though these typically imposed anchoring boundary conditions on the nematic order parameter, which are not appropriate to describe suspensions of swimmers such as bacteria.

As in the case of the disk, the transition to spontaneous flow can be understood as a linear instability of the equilibrium base state (phase I) as analyzed more precisely in Fig.~\ref{fig:annulusstability}. Here again, an infinite series of unstable modes exists, which involve increasingly complex azimuthal flow fields with alternating layers rotating in opposite directions, for which we show the marginal stability curves and profiles of the net particle velocity in Fig.~\ref{fig:annulusstability}(a)-(b). The first unstable mode, which causes the strongest pumping, is typically observed in simulations, though higher modes are also seen on rare occasions. As shown in Fig.~\ref{fig:annulusstability}(b), upstream swimming generally occurs near the annulus boundaries; this is always true of the outer boundary, though it is in some cases suppressed near the inner boundary when accumulation does not occur there as explained in Fig.~\ref{fig:basestateannulus}. The dependence of the transition to net pumping on the inner radius $r_\mathrm{min}$ is illustrated in Fig.~\ref{fig:annulusstability}(c): the marginal stability curve varies only weakly with $r_\mathrm{min}$ and has a non-monotonic dependence, which again reflects the change in the nature of the equilibrium base state found in Fig.~\ref{fig:basestateannulus}. In very large annuli ($r_{\mathrm{min}}\rightarrow \infty$), the effect of boundary curvature becomes negligible locally and the marginal stability curve asymptotes to that for a straight channel. 

\begin{figure}[b]
\centering
\includegraphics[scale=0.3]{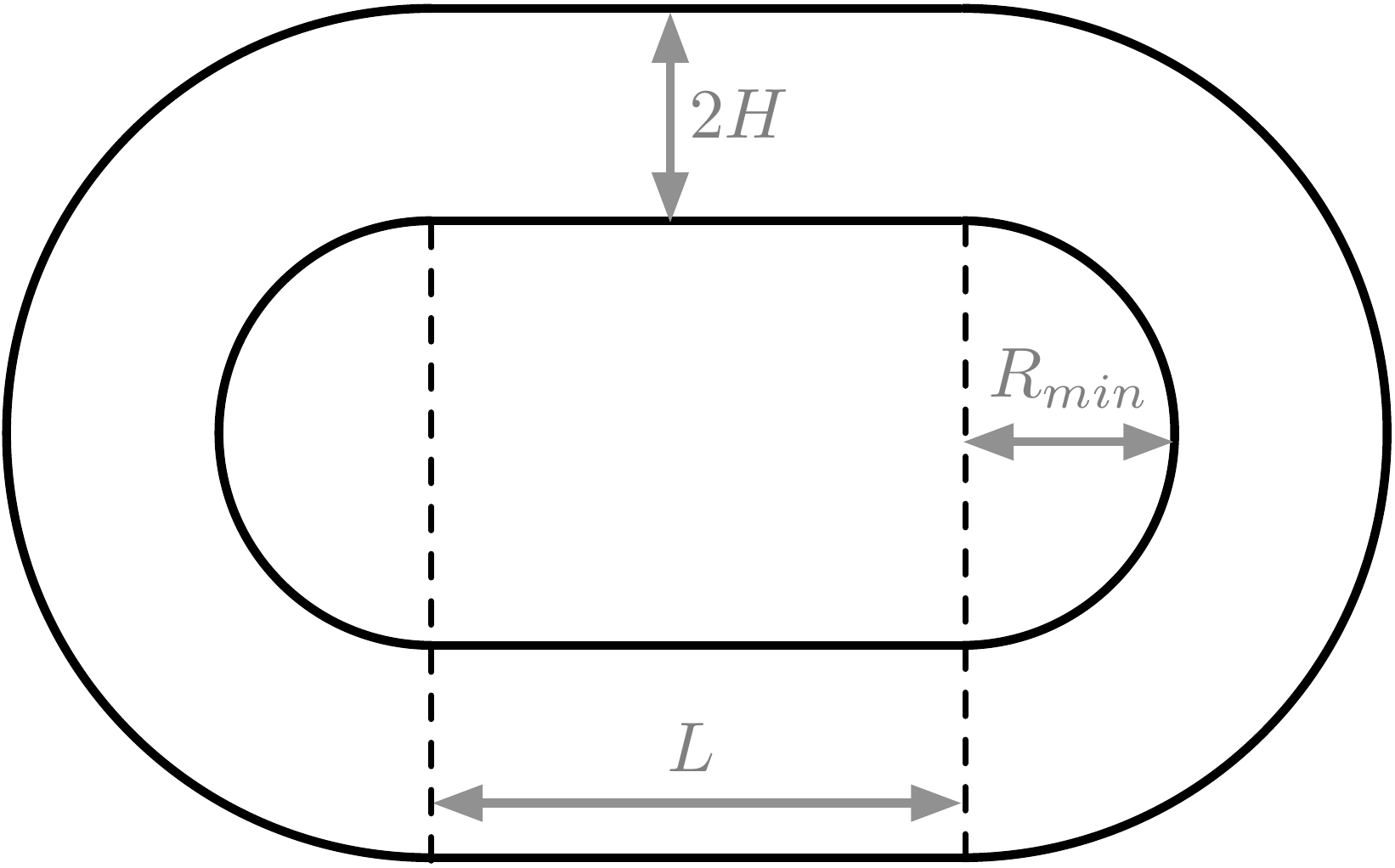}
\caption{Racetrack geometry: each boundary is composed of two straight sections (length $L$) and two half-circles (radii $r_\mathrm{min}$ and $r_\mathrm{min}+2H$). The distance between the two walls is $2H$, where the half-width $H$ is chosen for non-dimensionalization.     }  
\label{fig:racegeom}
\end{figure}

\begin{figure*}[t]
\centering
\includegraphics[scale=0.55]{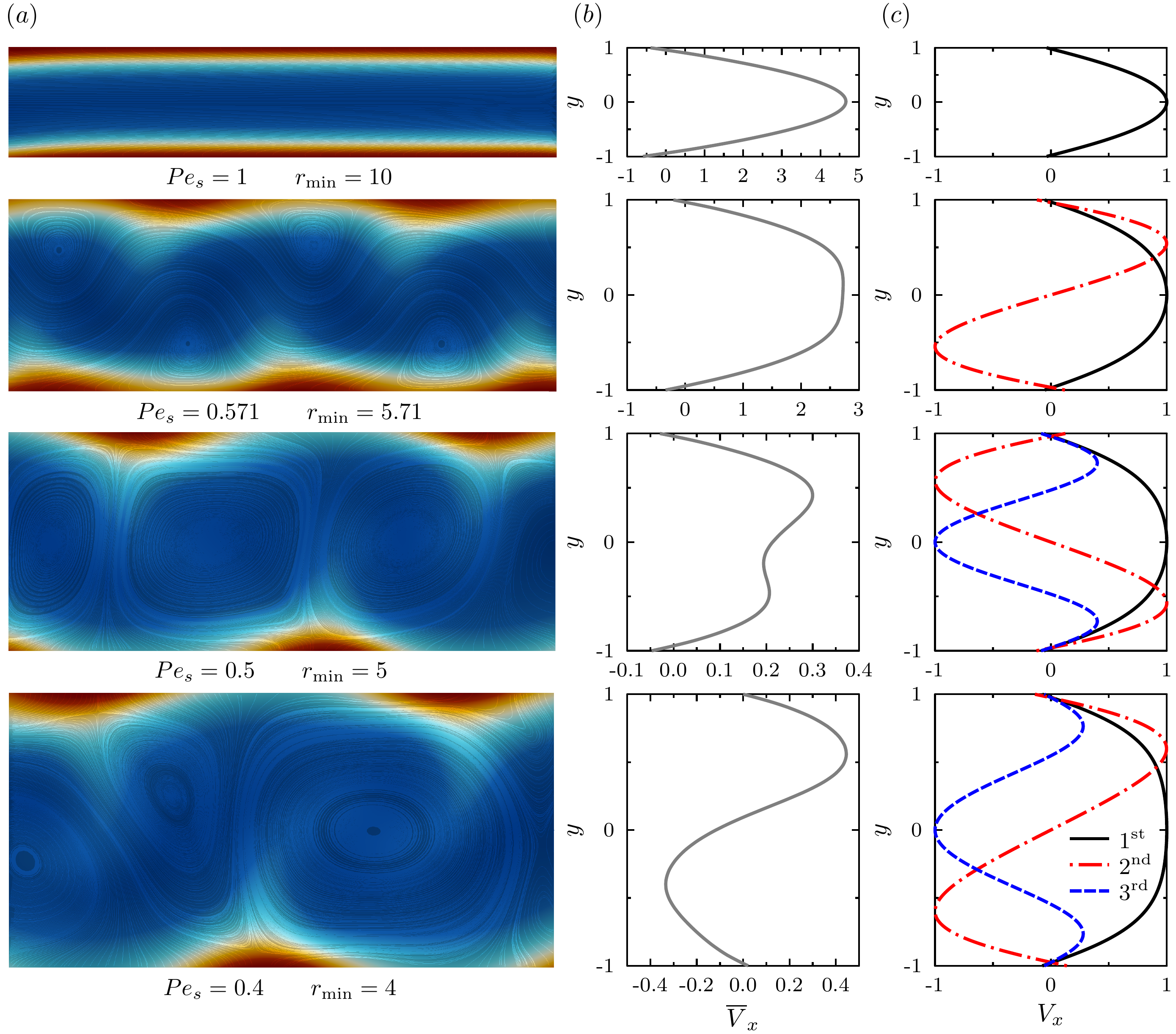}
\caption{(a) Concentration profiles and streamlines of the net particle velocity in straight sections of periodic racetracks, showing four different flow regimes (from top to bottom). (b) Average net velocity profiles across the gap in each case. (c) Unstable linear eigenmodes for these parameter values, obtained in straight channels. Results shown are for $\alpha=-60$ and $\Lambda=0.5$.  Also see electronic supplementary material for a movie showing the dynamics in each case.  }  
\label{fig:racetrack}
\end{figure*}

\vspace{0.3cm}

\noindent \textit{Phase III: Spontaneous flow with traveling waves. ---} At yet higher levels of activity, net pumping persists but the azimuthal symmetry of the flow is lost and periodic traveling density waves appear as shown in Fig.~\ref{fig:annulus}(c) as well as Fig.~\ref{fig:racetrack} below. Such waves were also observed in experiments on both bacterial \cite{Wioland16} and sperm suspensions \cite{Creppy15}. As illustrated in Fig.~\ref{fig:phasediagannulus}(b), the average azimuthal velocity $|\overline{V}_\theta|$ in this regime systematically decreases with activity level, as more and more of the motion takes place in the radial direction. As for the other steady states discussed previously, this one was found to be stable over long time intervals. A full characterization of these waves is beyond the scope of the present work. As either $|\alpha|$ is increased or $Pe_s$ decreased, the waves become more intense up to the point when the chaotic state of phase IV emerges; this state is similar to phase III observed above in circular domains.

\subsection{Periodic channels: racetracks}

For direct comparison with the experiments of Wioland \textit{et al.} \cite{Wioland16}, we consider as a final example racetrack geometries composed of straight sections of length $L$ connected by two half-annuli as shown in Fig.~\ref{fig:racegeom}. The characteristic length scale for non-dimensionalization is still taken to be the channel half-width $H$. We focus here on the dynamics in the straight sections; the various flow regimes in this case echo those found in circular annuli and are illustrated in Fig.~\ref{fig:racetrack}(a)-(b), showing instantaneous flow patterns and corresponding mean velocity profiles.  A transition to net pumping is first observed upon increasing channel width at a fixed value of $\alpha$, followed by the appearance of traveling waves. As the waves become stronger the flow takes the form of alternating counter-rotating vortices and eventually destabilizes into chaos. This phenomenology is identical to that observed in the experiments \cite{Wioland16} (see Fig.~4 of that reference). The mean velocity profiles are also consistent with the theoretically predicted unstable linear eigenmodes shown in Fig.~\ref{fig:racetrack}(c) for the same parameter values.  
 The last row in Fig.~\ref{fig:racetrack} demonstrates the possibility of a more complex flow regime that is only rarely observed: here the flow has a complex unsteady structure, but the mean velocity profile highlights two counter-flowing streams near the top and bottom walls and resembles the second unstable linear mode.

\begin{figure}[t]
\centering
\includegraphics[scale=0.9]{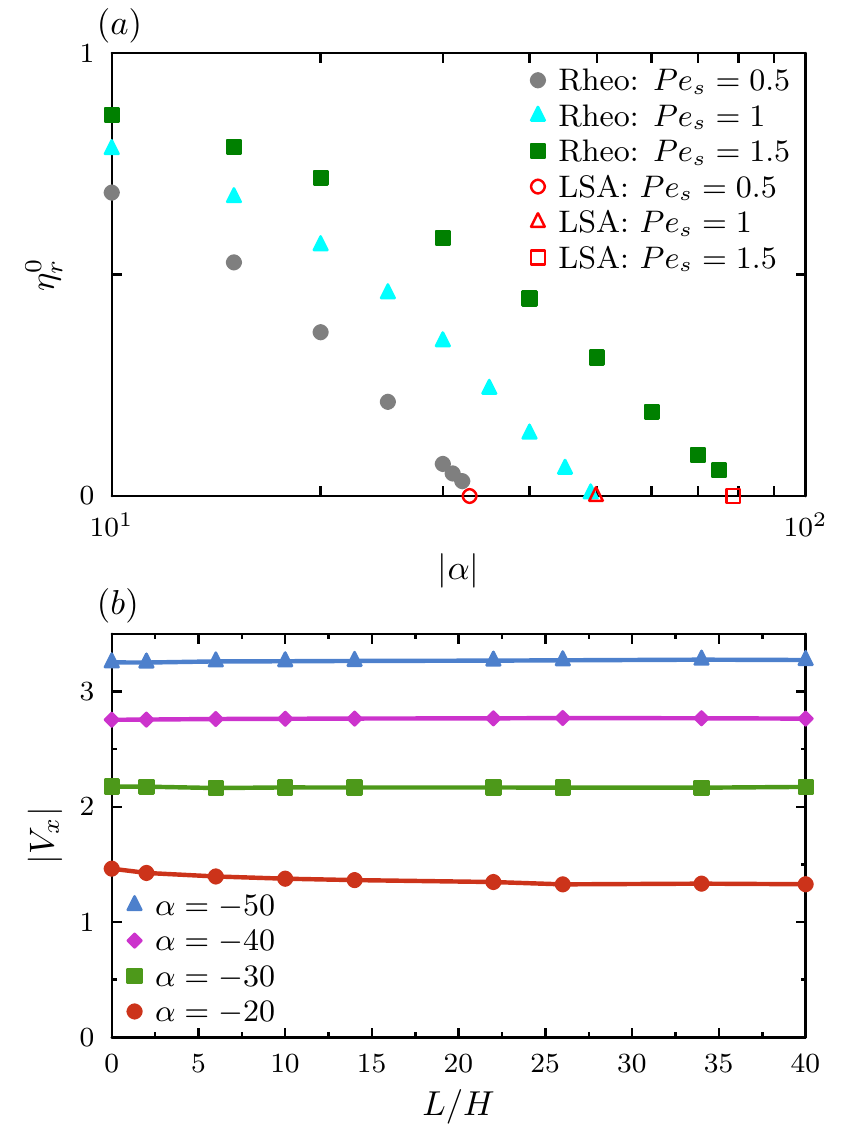}
\caption{(a) Zero-flow-rate relative velocity $\eta_r$ as a function of activity $|\alpha|$ in a suspension of pushers confined between two flat plates, obtained using the model of Alonso-Matilla \textit{et al.} \cite{Alonso16} As activity increases, the relative viscosity decreases and reaches superfluidity at the critical value of $|\alpha|$ for the onset of spontaneous flows as predicted by our linear stability analysis (LSA). This calculation was performed in three dimensions for ease of comparison with the rheological model. (b) Average longitudinal velocity $|V_{\parallel}|$ as a function of channel length $L$ in periodic racetracks in the spontaneous flow regimes (phases II and III).}  
\label{fig:rheology}
\end{figure}

We finish by describing the relationship between the onset of spontaneous pumping in confinement and the effective rheology of the system. In recent work, Alonso-Matilla \textit{et al.} \cite{Alonso16} calculated the effective relative viscosity $\eta_r$ in a dilute active suspension confined in a planar channel and subject to an applied pressure-driven flow, where $\eta_r$ is defined as the ratio of the flow rate in pure fluid by that in the presence of swimmers at a given pressure gradient. In agreement with previous experiments in the same geometry \cite{Gachelin13}, suspensions of pushers were found to enhance the flow, i.e.~decrease the effective viscosity of the suspension as a result of activity. This effect was found to be the strongest in the limit of vanishing flow strength, and the zero-flow-rate viscosity $\eta_{r}^0$ was further found to decrease with increasing $|\alpha|$, until it eventually reaches zero suggesting an apparent transition to superfluidity. The exact dependence of $\eta_r^0$ with activity is plotted in Fig.~\ref{fig:rheology}(a), where it is found that the value of $|\alpha|$ for which $\eta_r^0$ reaches zero coincides precisely with the prediction of the linear theory for the marginal stability of the equilibrium state and onset of spontaneous flow. This provides an additional interpretation for the transition to pumping as a consequence of apparent superfluidity: as the resistance of the system to flow is effectively zero, a small perturbation in the fluid velocity can amplify at no cost leading to unidirectional flow. In the flowing state, the input of mechanical energy by the swimmers exactly balances viscous dissipation in the solvent. This interpretation is further borne out by Fig.~\ref{fig:rheology}(b), which plots the mean flow rate in the spontaneous flow regime as a function of the length $L$ of the straight section of the racetrack. Remarkably, we observe that the flow rate in the pumping regime is completely independent of channel length, another signature of an effectively frictionless flow.

\section{Concluding remarks}

We have used a combination of numerical simulations and  theory to explore the effect of confinement and geometry on the onset and structure of spontaneous flows in semi-dilute suspensions of active swimming microorganisms. A mean-field theory based on the coupled Smoluchowski and Navier-Stokes equations was used to describe the dynamical evolution of swimmer configurations. We solved these governing equations numerically in two dimensions and compared our numerical results with predictions from a linear stability analysis. Our results agree well with prior experimental studies and were able to capture and explain the spontaneous directed motions arising in pusher suspensions. 

We first analyzed the dynamics of swimmers in circular disks, where three distinct flow regimes were found depending on the level of activity and degree of confinement: equilibrium with no   flow, a steady double vortex, and a swirling chaotic state. The equilibrium state manifests at low levels of activity or strong confinement, where the spatial and orientational distribution of particles is axisymmetric and the net disturbance flow generated by the swimmers vanishes. Particles accumulate at the boundaries and on average are radially polarized towards the wall, reaching their maximum polarization at the boundaries. Increasing the level of activity or decreasing confinement destabilizes the system into the double vortex state, and a mechanism based on the shear alignment of the swimmers in the disturbance flow they generate inside the accumulation layer was uncovered. By the same shear alignment mechanism, swimmers were in fact shown to orient against the flow throughout the domain allowing them to swim upstream near the boundary, thus leading to the double vortex structure reported in experiments where bacterial velocities were measured.  A further increase in the level of activity or a decrease in confinement originates a second transition to a turbulent-like chaotic state analogous to that observed in bulk systems.

We then turned our focus to swimmer dynamics in periodic geometries. Our simulations in circular annuli captured four flow regimes quite similar to those found in circular disks: an equilibrium state with no flow, an axisymmetric state with net fluid pumping, the emergence of traveling density waves, followed by a chaotic state. The transitions between regimes were again found to be governed by the level of activity and degree of confinement, with only a weak dependence on the inner radius dimension. Similar transitions were also observed in periodic racetracks, where we were able to connect the onset of spontaneous pumping with the effective rheology of the suspension. Specifically, the transition to net pumping was shown to occur at the same level of activity at which the zero-shear-rate viscosity becomes zero in a pressure-driven planar channel flow, suggesting that spontaneous flows in confinement are in fact a consequence of the apparent superfluidity of the system. This conclusion was further supported by observing that the net flow rate is independent of channel length in periodic racetracks. 

Our numerical and theoretical results both underscore the subtle interplay between confinement, geometry, and activity in semi-dilute active suspensions. While this study focused on fairly simple geometries previously considered in experiments, we anticipate that a wealth of more complex flow regimes might arise in other types of geometries. Continuum modeling as performed in this work proves to be a valuable tool for the understanding and prediction of these flows and could also play a useful role in the design of microfluidic devices for bioengineering applications involving bacteria.

\subsection*{Acknowledgments}

The authors thank Barath Ezhilan, J\'er\'emie Palacci, Hugo Wioland, Zvonimir Dogic, and Michael Shelley for valuable discussions. Funding from NSF Grants CBET-1532652  and   DMS-1463965 is gratefully acknowledged.

\section*{Appendix A: Analytical solutions for equilibrium states}

If the nematic order tensor is neglected $(\mathbf{D}=\mathbf{0})$, which is a good approximation at equilibrium as shown by the full numerical solution, simple closed-form analytical solutions can be derived for the radial concentration and polarization profiles which we provide here. In axisymmetric geometries, the steady governing equations for $c(r)$ and $m_r(r)$ obtained by setting the $\mathbf{F}_{c}$ and $\mathbf{F}_{m}$ in Eqs.~(\ref{eq:Fc})--(\ref{eq:Fm}) in the absence of flow simplify to: 
\begin{align}
-\frac{d}{dr}(rm_r)+2\Lambda Pe_s \frac{d}{dr}\left(r\frac{dc}{dr}\right)=0, \label{eq:ctheory} \\
-\frac{dc}{dr}+4\Lambda Pe_s \frac{d}{dr}\left[\frac{1}{r}\frac{d}{dr}(rm_r)\right]-m_r=0. \label{eq:mtheory} 
\end{align}
In a circular disk, the boundary conditions at $r=1$ are: 
\begin{align}
-m_r+2\Lambda Pe_s \frac{dc}{dr}=0, \label{eq:BCcr} \\
-\frac{c}{2}+2\Lambda Pe_s \frac{dm_r}{dr}=0, \label{eq:BCmr} 
\end{align}
and we also require that the solution remain bounded at $r=0$ and satisfy the normalization 
\begin{equation}
\int_{0}^{1}c(r) r dr = \frac{1}{2}.  
\end{equation}
Integrating Eq.~(\ref{eq:ctheory}) once easily shows that the boundary condition of Eq.~(\ref{eq:BCcr}) in fact applies everywhere across the gap and expresses the local balance between swimming and diffusive fluxes. Inserting Eq.~(\ref{eq:BCcr}) into Eq.~(\ref{eq:mtheory}) and manipulating then provides a modified Bessel equation for $m_r(r)$:
\begin{equation}
r^2\frac{d^2 m_r}{dr^2}+r\frac{dm_r}{dr}-(1+\Omega^2 r^2)m_r=0,
\end{equation} 
where
\begin{equation}
\Omega^2=\frac{1}{4Pe_s^2 \Lambda}\left(1+\frac{1}{2\Lambda}\right). 
\end{equation} 
After applying boundary conditions, the concentration and polarization profiles are obtained as
\begin{align}
c(r) = A_1 + A_2 I_0(\Omega r),  \quad 
m_r(r) = A_2 I_1(\Omega r), \label{eq:disksol}  
\end{align}
where the constants $A_1$ and $A_2$ are expressed in terms of incomplete Bessel functions as
\begin{align}
A_1&=1-\frac{2}{\Omega}I_1(\Omega)A_2, \\
A_2&=\left[(4\Lambda^2 Pe_s^2-1)I_0(\Omega)+\frac{2}{\Omega}I_1(\Omega)+4\Lambda^2 Pe_s^2\Omega^2 I_2(\Omega)\right]^{-1}.
\end{align}
The expression for $c(r)$ in Eq.~(\ref{eq:disksol}) is identical to that previously obtained by Yan \& Brady \cite{Yan15}. The solution inside an annulus is also easily obtained by applying boundary conditions of Eqs.~(\ref{eq:BCcr})--(\ref{eq:BCmr}) at both walls $r=r_\textrm{min}$ and $r_\mathrm{min}+2$ but is omitted here for brevity. The solution in a straight channel was also previously calculated by Ezhilan \& Saintillan \cite{Ezhilan15}.

\section*{Appendix B: Low- and high-$Pe_s$ marginal stability limits in a circular disk}

The stability analysis is performed by perturbing the governing equations about the equlibrium state, which we denote by $(c^0,\mathbf{m}^0,\mathbf{D}^0)$. We focus here on the marginal stability, for which the growth rate is set to zero. In large domains, the effect of boundary curvature on the structure of the accumulation layer is negligible, which prompts us to use Cartesian coordinates. Upon linearization of the equations, we arrive at a coupled system for the variables $m'_x$ and $D'_{xy}$ (where $'$ refers to perturbations): \vspace{-0.15cm}
\begin{align}
&0=-2 Pe_s \frac{dD'_{xy}}{dy}+4\Lambda Pe_s^2 \frac{d^2 m'_x}{dy^2}-\frac{3}{4}\alpha_c D'_{xy}m^0_y-m'_x, \\
&0=-\frac{1}{2}Pe_s\frac{dm'_x}{dy}+4\Lambda Pe^2_s \frac{d^2 D'_{xy}}{dy^2}-\alpha_c D'_{xy}\left(\frac{c^0}{4}+D^0_{yy}\right)-4D'_{xy}. \label{eq:evpb} v
\end{align}
These constitute an eigenvalue problem for $(m'_x, D'_{xy})$ with corresponding eigenvalue $\alpha_c$. 

In the low-$Pe_s$ limit, the dominant balance in Eq.~(\ref{eq:evpb}) is between the last two terms, which capture flow alignment and rotational diffusion. An estimate for $\alpha_c$ can then be obtained by balancing these two terms and by using the maximum values of $c^0$ and $D^0_{yy}$, which are attained at the walls: \vspace{-0.15cm}
\begin{equation}
\alpha_{c}^0\approx \frac{-4}{\frac{c^0_{wall}}{4}+D^0_{yy,wall}}=-\frac{32\Lambda}{1+2\Lambda}\qquad \mbox{as}\quad Pe_s\rightarrow 0, \vspace{-0.15cm}
\end{equation}
which indeed agrees with the full numerical solution of the eigenvalue problem as shown in Fig.~\ref{fig:stabilitydisk2}(a). 

In the high-$Pe_s$ limit (strong confinement), the dominant balance now takes place between the second and third terms in Eq.~(\ref{eq:evpb}), which describe translational diffusion and shear alignment. While the use of Cartesian coordinates is no longer justified in this case, the form of the equations still suggests a scaling of the type \vspace{-0.15cm}
\begin{equation}
\alpha^\infty_c\approx -\beta \Lambda Pe_s^2 \qquad \mbox{as}\quad Pe_s\rightarrow \infty, \vspace{-0.15cm}
\end{equation}
which is again supported by numerical data with $\beta\approx 150$ as shown in Fig.~\ref{fig:stabilitydisk2}(b).

\bibliography{rsc}

\end{document}